# Precise characterization of a silicon carbide waveguide fiber interface

Marcel Krumrein[1], Raphael Nold[1], Flavie Davidson-Marquis[2,3], Arthur Bourama[1], Lukas Niechziol[1], Timo Steidl[1], Ruoming Peng[1], Jonathan Körber[1], Rainer Stöhr[1], Nils Gross[4], Jurgen Smet[4], Jawad Ul-Hassan[5], Péter Udvarhelyi[6,7,8], Adam Gali[6,7,8], Florian Kaiser[1,2,3], and Jörg Wrachtrup[1,9]

1     3rd Institute of Physics, IQST, and Research Centre SCoPE, University of Stuttgart, Stuttgart, Germany.

2     Materials Research and Technology (MRT) Department, Luxembourg Institute of Science and Technology, Belvaux, Luxembourg.

3     Department of Physics and Materials Science, University of Luxembourg, Belvaux, Luxembourg.

4     Solid State Nanophysics, Max Planck Institute for Solid State Research, Stuttgart, Germany.

5     Department of Physics, Chemistry and Biology, Linköping University, Linköping, Sweden.

6     Wigner Research Centre for Physics, Budapest, Hungary.

7     Institute of Physics, Department of Atomic Physics, Budapest University of Technology and Economics, Budapest, Hungary.

8     MTA-WFK Lendület "Momentum" Semiconductor Nanostructres Research Group.

9     Max Planck Institute for Solid State Research, Stuttgart, Germany.

## 1. Abstract

Emitters in high refractive index materials like 4H-SiC suffer from reduced detection of photons because of losses caused by total internal reflection. Thus, integration into efficient nanophotonic structures which couple the emission of photons to a well-defined waveguide mode can significantly enhance the photon detection efficiency. In addition, interfacing this waveguide to a classical fiber network is of similar importance to detect the photons and perform experiments. Here, we show a waveguide fiber interface in SiC. By careful measurements we determine efficiencies exceeding 93 % for the transfer of photons from SiC nanobeams to fibers. We use this interface to create a bright single photon source based on waveguide-integrated V2 defects in 4H-SiC and achieve an overall photon count rate of 181 kilo-counts per second. We observe and quantify the strain induced shift of the ground state spin states and demonstrate coherent control of the electron spin with a coherence time of $T_2 = 42.5$ µs.

## 2. Introduction

Silicon carbide (SiC) is an emerging host material for quantum emitters. Especially the silicon monovacancies V1 and V2 in 4H-SiC attracted attention as spin quantum bits in an industrial scale semiconductor material. Both vacancies possess encouraging spin-optical properties, such as long coherence times[1,2], accessibility to a nuclear spin bath with two different atomic species[3,4] and lifetime-limited emission lines for temperatures up to 20 K[5]. However, due to the presence of a non-radiative, long-lived metastable state[6], their brightness is low, such that phenomena like non-destructive spin quantum state readout become challenging. Moreover, low collection efficiency resulting from total internal reflection in bulk material further contributes to the reduced brightness. Integrating those defects into nanophotonic structures, such as solid immersion lenses and nanopillars, decreases the emission angle and, thus, increases the collection efficiency[7-9]. The photon count rate can be further increased with SiC nanobeams. Previous work[10] demonstrated that waveguide-integrated V2 defects maintain their pristine spin-optical properties. However, those waveguides lack an efficient interface to collect the emitted photons.

Several approaches were proposed and realized to extract photons from nanobeam structures: grating couplers with efficiencies up to 81 %[11–13], lensed fibers with 69 %[14–16], 3D tapered waveguide-to-fiber couplers with 85 %[17], circular Bragg gratings with 63 %[18], and tapered fibers with 93 %[19–21]. Because of the highest reported coupling efficiencies, tapered fibers offer the most promising avenue. Hence, their adoption to silicon carbide nanobeams seems very promising.

In this work, we demonstrate an efficient waveguide-fiber interface in 4H-SiC at ambient conditions with efficiencies exceeding 93 %. This interface is tested for its robustness and dependencies on various parameters, such as fiber and taper geometry, wavelength, and overlap length. We support these results by extensive finite difference time domain (FDTD) simulations. Silicon vacancies were implanted into these waveguides creating a bright single photon source at room temperature. Finally, we perform Rabi and Hahn-echo measurements to show the accessibility of the spin and a speed-up of the measurement time compared to bulk emitters.

## 3. Main text

To fabricate single-mode waveguides (fig. 1a), nanobeams were patterned into a positive-tone electron beam resist via electron beam lithography. After resist development, nickel was deposited, followed by the lift-off process. The resulting metal mask was first vertically transferred into SiC via reactive ion etching using an $SF_6$ plasma, and then, etched into triangular shape by usage of a graphite Faraday cage with an undercut angle of 36°. Finally, the metal mask was removed in nitric acid and Piranha solution. We obtain freestanding waveguides with a triangular cross-section and flexible waveguide length and width. To hold the nanobeam, support structures as in ref. [20] were used that are 40 µm apart from each other and have a length of about 5µm. Additionally, the waveguide width and, thus, also the waveguide height, is reduced linearly to zero on both ends (see fig 1b) forming tapers that are used for in- and outcoupling of light. Overlapping those with tapered optical fibers forms an efficient interface to a classical fiber network, see fig. 1d. The power transfer can be understood in the following way:

By decreasing the waveguide width and, thus, the cross-section, the light mode remains confined in a fundamental eigenmode of the nanobeam whereas higher order modes are suppressed. Additionally, this guided eigenmode is getting closer to the waveguide edge facilitating the coupling to objects close to the edge. By approaching the overlap region with the tapered optical fiber, hybridized supermodes can be formed to transfer power from the higher-index material SiC into the lower-index material SiO2 provided that this change is slow, and the transfer can happen in an adiabatic manner. In this context, adiabaticity means that the overlap region is larger than the wavelength. Because of time reversal symmetry in Maxwell's equations, this mode transfer shows the same characteristics and efficiencies as that of transferring light from the optical fiber into the waveguide.

The tapering procedure of the optical fibers[22] is described in the methods section. It creates highly reproducible conical fiber tips. However, after the etching process, the fiber-tip is fragile and prone to breakage, as visible in the inset of fig. 1c. This can lead to reduced efficiency in the transmission of light.

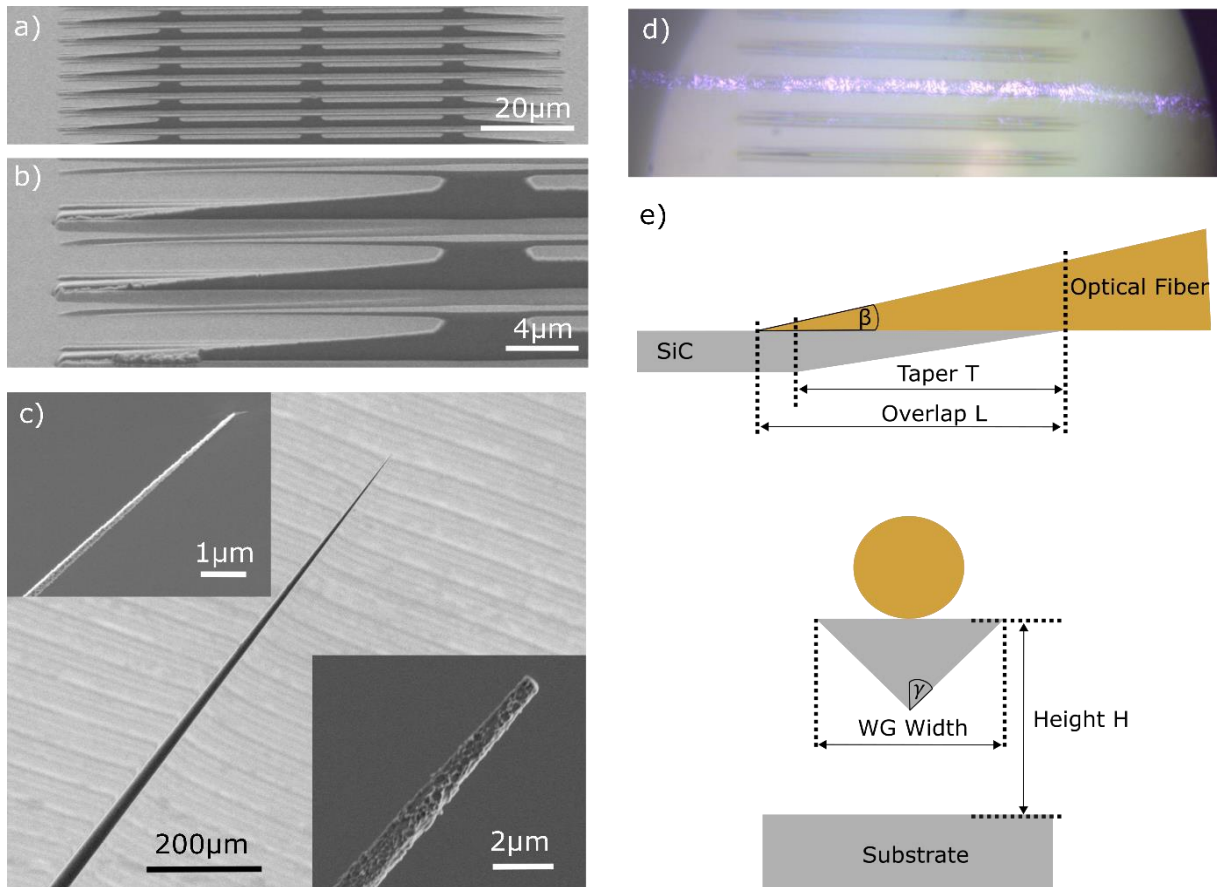

**Figure 1:** Geometries and working principle of the waveguide-fiber interface. (a) SEM image of fabricated waveguides with a zoom-in image on the waveguide tapers in (b). (c) SEM image of a tapered optical fiber (1060XP) with a full-opening angle of about 1.95°. The inset on the top left shows the close-up of an optimal fiber tip, whereas the fiber tip in the inset on the bottom right is broken with a tip radius of about 250 nm. (d) Microscope image of two tapered optical fibers attached to the waveguide from the left and right side. The lighting of the waveguides indicates that some laser light is propagating through the waveguide. (e) Schematic of the waveguide-fiber interface looking from the side (top image) and a cross-section along the propagation direction (bottom image) with all relevant, geometric parameters.

We performed FDTD simulations with a commercial software package[23] to find the optimal waveguide geometry for hosting the silicon monovacancy V2. The half-opening angle at the apex of the waveguide is determined by the Faraday cage and fixed to $\gamma \sim 36°$. For a nanobeam width of 490 nm, the transmission of light propagating at 960 nm (center of the PSB of V2) is single-mode and close to its maximum. Hence, we chose a waveguide width of 490 nm for all fabricated devices presented in this work. Additional information can be found in supplementary note 1.

In previous publications, the waveguide-fiber efficiency was determined by reflection measurements in cavity structures. This comes along with the difficulty of properly distinguishing between the mode transfer itself and cavity reflection. In contrast, in this work we characterize the interface by attaching two tapered optical fibers to both ends of the waveguide, and thus, measuring the transmitted power. A microscopy image of this measurement configuration is given in fig. 1d. Fig. 2e confirms the time reversal symmetry, showing that coupling into the waveguide (in-coupling) follows the same behavior as into the fiber (out-coupling). The waveguide-fiber interface efficiency $\eta_{\mathrm{WFI}}$ can be calculated by the equation $\eta_{\mathrm{trans}} = \eta_{\mathrm{WFI}}^2 \cdot \eta_{\mathrm{coupler}} \cdot \eta_{\mathrm{wg}}$, with the transmitted power efficiency $\eta_{\mathrm{trans}}$, the fiber connector efficiency $\eta_{\text{coupler}} \approx 0.9$ and the waveguide transmission efficiency $\eta_{\text{wg}} \approx 0.98$. In figs. 2b-d, we do not account for the additional losses $\eta_{\text{coupler}}$ and $\eta_{\text{wg}}$. Therefore, the quoted values are lower limits and the actual $\eta_{\text{WFI}}$ is even larger.

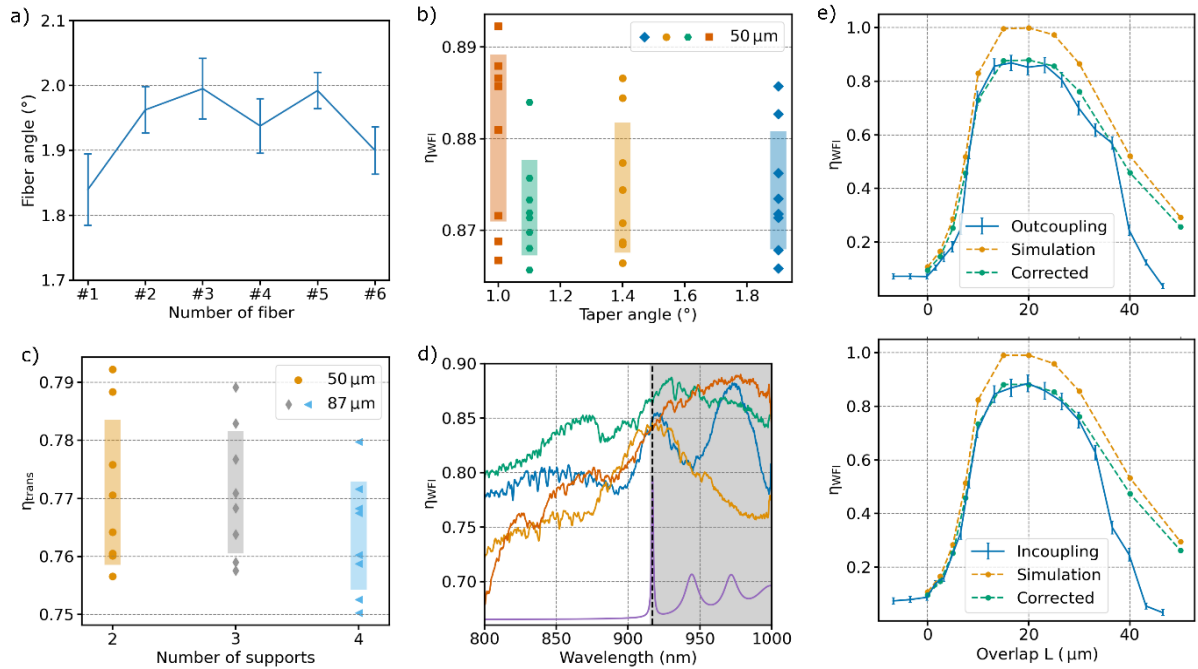

**Figure 2:** Performance of the waveguide-fiber interface. (a) Full-opening angle of six, arbitrarily chosen, tapered fibers showing the high reproducibility of our etching technique. (b) Single-sided coupling efficiency for varying taper angle. (c) Transmission for varying number of support structures of the waveguide. The rectangles in b and c indicate the standard deviation of the data. (d) Wavelength-dependency of the interface for four different waveguides. The purple curve shows the theoretical emission spectrum of V2 at cryogenic temperatures and the grey area indicates the PSB intended to collect from the emitter. The dashed black line marks the wavelength used for the characterization measurements in b, c, and e. (e) Single-sided coupling efficiency for different overlap lengths L. The upper image represents the case of a changing overlap on the out-coupling side, whereas the lower image covers the overlap on the in-coupling side. The orange data are simulation results for the same geometry with a broken tip radius of 170 nm. Correcting those data for additional losses results in the green curve.

First, the geometry of the tapered optical fiber was investigated. The shape of the fiber is determined by the etching process which produces highly reproducible tapered fibers with an opening angle of around 1.95°, see fig. 2a. However, our simulation revealed that, as long as the fiber angle is below 4°, no significant change in the coupling performance is observed (see supplementary note 2). Likewise, fig. 2b shows the efficiency relative to the waveguide taper angle. For values ranging from 1 to 2°, maximum efficiencies between 86.9 % and 89.2 % are obtained with mean values between 87.6 % and 88.4 %. We conclude that the changes of the taper angle contribute less than 1 % to the overall maximum efficiency. Taking $\eta_c$ and $\eta_{wg}$ into account, $\eta_{\text{WFI}}$ exceeds 93 %. Hence, for waveguide tapers below 2° and fiber tapers below 4°, adiabaticity of mode transfer is realized. Those values are very similar to previously reported values on other platforms like diamond with reported efficiencies around 93 %. We also estimate the losses induced by propagation through the waveguide. For this, we fabricated waveguides with different lengths and number of support structures. For nanobeams with three supports (grey diamonds in fig. 2c), the average transmission is (76.4±0.9) % and for four supports (blue triangles), we measure (77.1±1.1) %. Accordingly, the losses induced by the support structures are at most 1 %. Their impact on the waveguide transmission comes through an abrupt change of the waveguide geometry, much faster than the adaption of the guided waveguide mode through adiabatic changes. However, since the amplitude of the mode at the support structures is small, their impact on losses is small as well.

Comparing waveguides with a varying total length of 50 µm (orange circles in fig. 2c) and 87 µm (grey diamonds) show the same transmission of (77.1±1.3) %. Hence, the propagation loss induced by the waveguide itself is beyond our measurement accuracy. In contrast, fig. 2d shows that the wavelength of the photons is a more critical parameter. The main reason is the geometrical optimization of the nanobeam to support modes at a wavelength of 960 nm. Therefore, large deviations towards shorter wavelengths cause an increase in loss due to the multimode nature of the guided field. The waveguide-fiber interface depends much weaker on the wavelength. However, for wavelengths between 917 nm (ZPL of V2) and 1000 nm (PSB of V2), the efficiency forms a plateau with values around 88 %. Taking the additional losses in the system into account, the isolated waveguide-fiber transmission efficiency is estimated to be 93-94 %. Accordingly, the fabricated waveguides possessing such a waveguide-fiber interface can be used for collecting the emission of the V2 center efficiently. As a last performance test, in fig. 2e we checked the importance of the overlap length L of the tapered optical fiber and the waveguide taper (as illustrated in fig. 1e). The coupling efficiency is maximized for an overlap of about 15 µm demonstrating the overall robustness of the system. Due to the symmetry of the in- and out-coupling systems, varying the fiber-waveguide contact length at either end of the waveguide is equivalent. This is observed experimentally as well. Additionally, we simulated the coupling efficiency as a function of overlap L between the tapered waveguide and the fiber, resulting in the orange data points in fig 2e. From the plateau width, the radius of the intentionally broken fiber used for this measurement is estimated to be 170 nm based on the comparison with simulation data. Assuming a total loss of 22 % of the entire system, originating from in-and out-coupling, support structures and optical elements, a correction can be applied to the simulation data (green data points). This yields excellent agreement with the experimental data up to an overlap length of about 30 µm. For larger L, the measured efficiency drops faster than expected from the simulations. They assume a physical contact between waveguide and fiber across the entire length L, but this is very difficult to achieve in experiments for large overlap lengths.

Next, we investigate waveguides with silicon vacancy centers created by electron irradiation. For details, see Methods. To investigate those structures, a confocal path was installed to excite the emitters from the top with an air objective of 0.75 NA and a pulsed laser operating at 780 nm. The fluorescence was collected via the waveguide-fiber interface. The entire setup is shown in fig. 3a. Confocal raster scans of the waveguides after removal of the nickel mask in nitric acid and Piranha solution showed a high background fluorescence from the waveguide surface. The waveguide length and emitter density were chosen such that we expected 3 to 4 V2 centers per waveguide. In contrast, about 60 fluorescent emitters were found in each waveguide. Contamination originating from ambient atmosphere, e.g., $H_2O$, and from the formation of carbon based organic molecules can be removed by desorption[24]. Hence, the sample was annealed at 600°C in argon atmosphere for 30 minutes. However, after the annealing procedure the surface fluorescence massively increased. EDX measurements revealed the nature of the contaminations to be metallic, such as nickel left over from the metal mask. By annealing, they diffuse into the substrate and form luminescent defects. Also, etching of the metal mask leads to the same formation of background defects. As most of them are located close to the surface, a post soft-ICP etch step with $SF_6$ plasma was applied to remove few nanometers of the surface reducing the disturbing photoluminescence significantly. After this step, about 30 well resolved fluorescing spots per device were found. Representative confocal raster scans of three different waveguides are displayed in figure 3b. Lifetime measurements revealed that the excited state lifetimes of those surface emitters are around 6 ns and, thus, shorter than that for the V2 centers of 9 ns. Hence, pulsed excitation and time gated detection of fluorescence achieves a better signal-to-noise ratio for the V2's compared to cw-excitation.

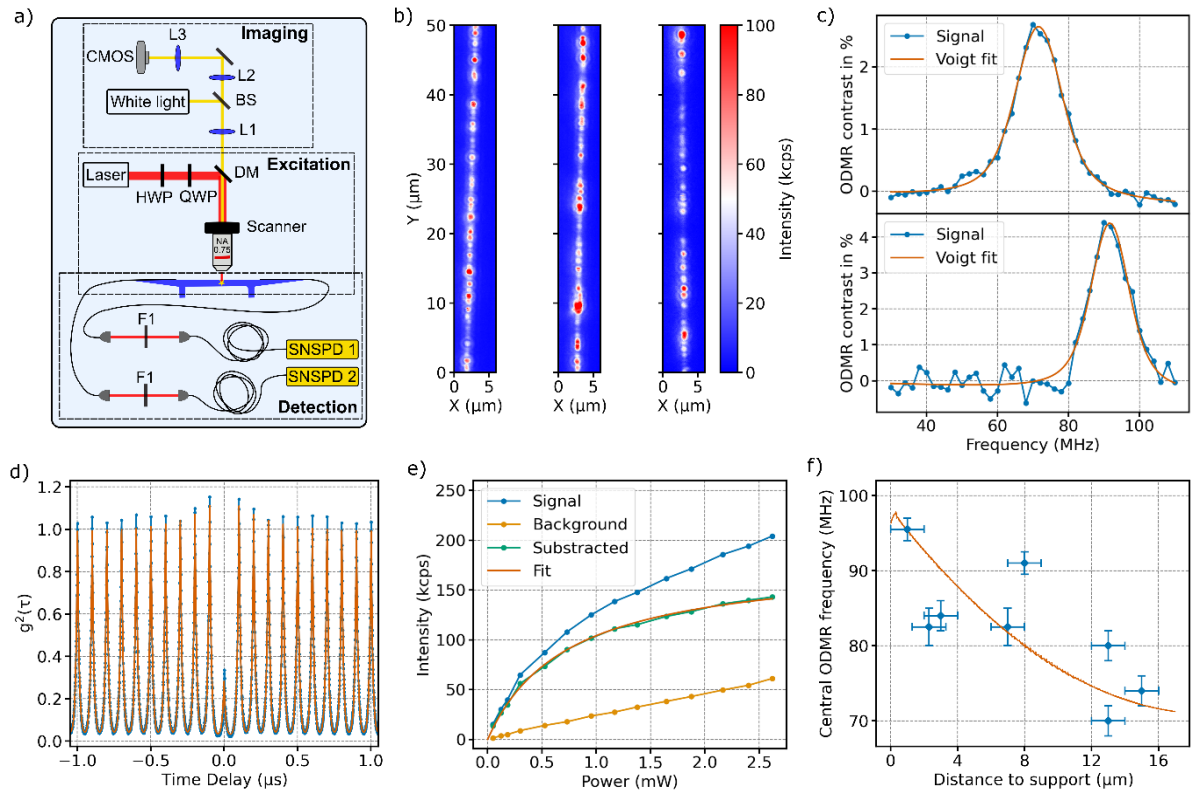

**Figure 3:** Characterization of the waveguide-integrated defects. (a) Setup of a home-built confocal microscope consisting of an imaging system, an excitation and detection path. For excitation, a pulsed 780 nm laser is focused on the waveguide from the top with the help of a free-space objective (with NA 0.75) after passing a half-wave plate (HWP), a quarter-wave plate (QWP), and a dichroic mirror (DM). The emitted photons are collected with two tapered 1060XP fibers and sent to superconducting nanowire single-photon detectors (SNSPD) after filtering of the exciting laser photons (F1). For simplified fiber positioning, an imaging system consisting of a white light source, three lenses (L1-L3) and a CMOS camera was installed. (b) Confocal raster scans of three different waveguides with a high concentration of defects, most of them attributed to surface defects (for more details, see text). (c) ODMR spectrum of a V2 center with its central peak position at $(71.6 \pm 0.11)$ MHz. (d) Auto-correlation measurement of the defect shown in (c) with an anti-bunching dip of $g^2(0) = 0.268 \pm 0.003$ proving the single photon emission character. No corrections to the data were applied. (e) Saturation study of the same defect with a background-corrected saturation intensity of $\sim 181$ kcps. (f) Dependency of the ODMR peak frequency on position in the waveguide. The red curve originates from a strain simulation along the waveguide.

The V2 center has a paramagnetic ground state with S=3/2. As a result, an external magnetic field splits the ground state into four sublevels. However, even in zero external magnetic field the non-averaged spin-spin interaction (zero field splitting: ZFS) splits the ±1/2 from the ±3/2 states. As the ZFS of the ground state (GS) at zero magnetic field is 2D=70

MHz, we expect an ODMR peak at this frequency position for V2 centers. However, the experimentally measured central peak positions varied between 70 and 95 MHz for different defects, a noteworthy deviation from bulk emitters of up to 25 MHz. In the following, we discuss possible explanations for this effect.

First, coupling to nearby nuclear spins causes a splitting of the ODMR peak that depends on the nuclear species and the position of those nuclei. But in such a case, two transitions should be observable in the spectrum unless the second peak is suppressed. This scenario can in principle happen for polarized emitters. However, it is very unlikely that 7 out of 8 emitters are coupled to nearby nuclear spins and at the same time polarized in one energy level, considering that the sample contains only a limited number of nuclei with a non-zero magnetic moment, 4.7% of Si29 and 1.16% of C13 atoms.

Second, an electric field originating from point defects (such as carbon vacancies) or foreign atoms (such as nitrogen atoms) in the local environment could shift the ground state spin levels via the Stark effect. But in twin samples from the same growth process, we haven't seen any effect of shifting the GS ZFS. This means that possible sources of electric fields must be located near the surface of the waveguide. A single point charge with 10 elementary charges at the waveguide surface would for instance cause an electric field strength of $|E|\sim 15\,\mathrm{kV/cm}$ at the center of the nanobeam where we expect our defects to be located. The calculated *ab initio* Stark-shift for the V2 center's GF ZFS is $\Delta D \approx 13\,\mathrm{Hz}\frac{\mathrm{cm}}{\mathrm{V}} \cdot E$, resulting in an expected ODMR shift of 195 kHz. This is orders of magnitude smaller than the observed deviation of 25 MHz and suggests that electric fields are a very unlikely reason for the observed ODMR shifts. Note that we neglected any other charge distribution as well as screening effects that may reduce the effective electric field strength, so that $|E|\sim 15\,\mathrm{kV/cm}$ is likely even an overestimated value.

Third, strain in the environment of color centers influences the GS ZFS. For diamond defects, this effect has already been investigated extensively. It offers an opportunity to tune the emitters when strain can be applied in a controlled fashion[25–27]. Similarly, large strain in the lattice of silicon carbide can change the GS ZFS of embedded color centers by tens of MHz[28,29]. The associated spin-phonon interaction is described by the Hamiltonian in cartesian coordinates[30,31]

$$H = \sum_{\alpha\beta} \varXi u_{\alpha\beta} S_\alpha S_\beta,$$

with the deformation potential constant $\varXi$, the spin-3/2 operator $\boldsymbol{S} = (S_x, S_y, S_z)$, and the deformation tensor $u_{\alpha\beta}$. This Hamiltonian can be decomposed in a parallel and a perpendicular strain component relative to the crystal c-axis, $\varepsilon_{\mathrm{para}}$ and $\varepsilon_{\mathrm{perp}}$ respectively. Accordingly, for color centers located in such strained environments, a shift of the energy levels dependent on the applied strain is expected, whereas Kramer's degeneracy stays preserved in the absence of a magnetic field. This leads to an increase of the GS ZFS and, thus, to a shift of the ODMR peak position. Indeed, our *ab initio* parameters are, $\varXi_{\mathrm{para}} = 2.8\,\mathrm{GHz/strain}$ and $\varXi_{\mathrm{perp}} = -1.9\,\mathrm{GHz/strain}$, respectively.

Because of the geometry of our waveguide structures, the strain along the nanobeam direction is expected to reach a maximum closest to the support structures and to decrease with increasing distance. A detailed analysis of our data reveals a correlation between the ODMR shift and the distance of the emitter to the next support structure as visible in fig. 2f. This behavior is confirmed by simulations with the *COMSOL Multiphysics software*[32] (red curve in fig. 2f). To check the strain inside the waveguides, we performed Raman spectroscopy measurements. From the shift of several phonon peaks in the Raman spectrum, it is possible to extract the strain in the nanobeam. We measure values of about 0.1 % for $\varepsilon_{\text{para}}$ and -0.35 % for $\varepsilon_{\text{perp}}$. With these strain values, we calculated the expected ODMR shift to 23.4 MHz matching the experimentally measured shifts very well. For more details, we refer to supplementary Note 5. In these measurements, the laser spot is about 0.5 µm large, thus, the obtained values for strain are average values over a larger volume. This is particularly noteworthy as the local strain environment of the defects depends strongly on the exact position in the waveguide cross-section. Given the fact that the expected shift of the GS ZSF extracted from the strain measurements agree with the experimental data rather accurately, we attribute the observed ODMR peak shifts to the significant amount of strain inside the nanobeam.

Due to the high density of interfering surface defects in the waveguides, most of the V2 centers do not show a clear antibunching behavior in second order auto-correlation measurements. Nonetheless, a V2 center with a clear single photon character as in fig. 3d can be found in every second waveguide. Saturation studies on these single V2s were performed and after correction of the linear background emission, a saturation fluorescence of 181 kcps was obtained, see fig. 2e. This is, to our knowledge, the highest count rate observed for the V2 center without making use of a significant Purcell enhancement. In supplementary note 7, a comparison with previously reported count rates in the literature is given. In our room temperature experiments with non-resonant excitation, the signal-to-noise ratio is found to be between 5 and 9 for single V2s (see supplementary note 8) and, slightly smaller than for bulk emitters. This is due to the presence of interfering surface defects in the waveguide. Upon resonant excitation this background will be significantly reduced, making waveguides a prime choice for efficient photon collection.

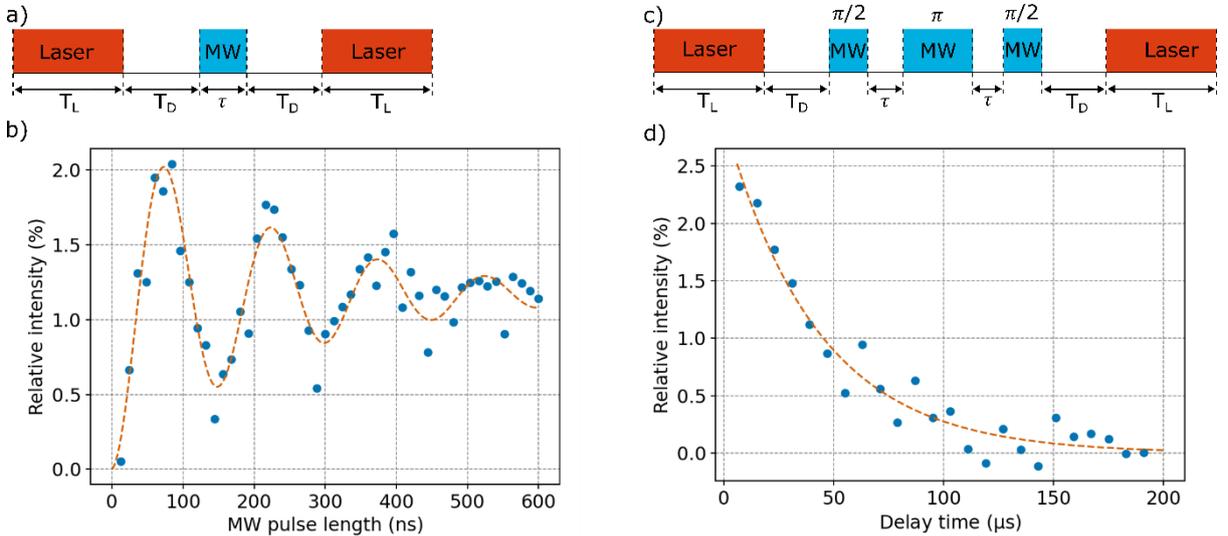

**Figure 4:** Coherent spin control. (a) Pulse scheme for measuring Rabi oscillations. (b) Rabi oscillations observed on a V2 emitter at room temperature and in absence of an external magnetic field. The Rabi frequency is $f_{\text{Rabi}} = 6.65$ MHz with a decay time of $\tau_{Rabi} = 234\ ns$. (c) Pulse sequence for detecting the Hahn-echo signal. (d) Hahn-echo decay signal of the same emitter as in (b). An exponential decay is fitted to the data revealing a coherence time of $T_2 = (42.5 \pm 5.3)$ µs.

Finally, we demonstrate coherent spin control of the waveguide-integrated emitters. First, Rabi measurements were performed to disclose the pulse length for π and π/2 pulses. To this end, the measurement scheme illustrated in fig. 4a was used. We observe a clear oscillation at a frequency of 6.65 MHz (fig. 4b). The Rabi signal decayed very fast with an exponential decay time of 234 ns. There are several explanations for this fast decay: (i) The spin environment is very noisy and leads to spin flip-flop events. (ii) A drift of the microwave may occur during the measurement time. This would lead to a mismatch between Rabi frequency and the driving microwave field. (iii) The strain within the waveguides can change due to instabilities of the environmental conditions, such as for instance temperature changes. Second, the coherence time was determined by performing the Hahn-Echo sequence in fig. 4c. Fig. 4d demonstrates an exponential decay of the spin signal with a fitted decay time of $T_2 = (42.5 \pm 5.3)$ µs. Considering that this measurement was run at ambient conditions in absence of any externally applied magnetic field, the obtained T2 time is comparably long for the V2 center[2,33,34]. However, the absence of a preservation of the spin state in the first part of the signal indicates a fast spin bath environment. This also explains the fast decay of the Rabi signal. The presence of nickel related defects in the nanobeam could be one source of spin-spin interaction leading to the fast dephasing. Hence, efforts to further reduce the density of those defects should be beneficial. Note that because of the high photon count rate of the emitter, the measurement times for Rabi and Hahn-Echo experiments were significantly shorter compared to defects located in bulk material.

## 4. Summary and Outlook

Here, we presented an interface connecting silicon carbide waveguides with an optical fiber system and showed efficiencies up to 93 %. Our approach of tapering both, the waveguide ends and the optical fibers, was already successfully demonstrated for several material platforms, such as silicon nitride[19], diamond[20,21,35] silicon[36,37], and lithium niobate[21]. What makes this technique particularly attractive is the possibility to simultaneously approach several structures on a fully packaged chip connecting several nanophotonic devices. The performance was already experimentally tested previously in the literature. Also, extensive simulations were run previously to identify the crucial parameters. Here, we have combined both: an exhaustive characterization of the interface, and we successfully demonstrated the applicability to the silicon carbide platform.

For diamond nanostructures, coupling efficiencies up to 93 % were reported in Ref. [19,21] which is comparable to the values reported here, also 93 %. The integration into cryogenic systems was demonstrated for the diamond platform[21,3,20], but remains to be shown for SiC. For silicon waveguides operated at a wavelength above 1400 nm, efficiencies around 80 % are possible[37] which is lower than for SiC nanobeams. Hence, we report the highest coupling efficiencies for telecom wavelengths beyond 900 nm.

Alternative coupling techniques[38], as cleaved fibers[39] and grating couplers[40], can provide comparably high coupling efficiencies. However, they either have restricted access only to waveguides located at the edge of the sample[39], or simultaneous access to several structures is not possible[40].

Of course, our platform is not only limited to host the silicon vacancy center V2. It is easily possible to adjust the width of the waveguide to also host the V1 center, as well as the divacancies PL5-PL7 in 4H-SiC. Additionally, this approach is not restricted to 4H-SiC, but can also be applied to other silicon carbide polytypes as 3C- and 6H-SiC.

# 5. Methods

## 5.1 Sample Preparation

For most measurements, a 4H-SiC a-plane sample was taken and overgrown with a 10 µm thick epilayer by chemical vapor deposition. The free carrier concentration is $7 \times 10^{13}$ cm$^{-3}$ with natural abundance of isotopes. For implantation, electron irradiation with an energy of 5 MeV and a dose of 2 kGy was used. Afterwards, the sample was annealed at 600°C in argon atmosphere for 30 min. For the measurement of strain in the waveguides, a commercially available a-plane 4H-SiC sample was used.

## 5.2 Waveguide Fabrication

For creating an etching mask, a 400 nm thick electron beam resist (AR-P 6200 from *Allresist*) was patterned with 20 kV electron beam lithography. A 200 nm thick nickel mask was applied with an electron beam evaporator. After lift-off process, the metal mask was transferred vertically into SiC with RIE using SF$_6$ plasma (100 W, 20 sccm, 7.5 mTorr). To suspend the nanobeams, a home-built Faraday cage of graphite was used resulting in a triangular cross-section. The support structures of the devices were 30 % larger than the nanobeam width. The etch mask was removed in nitric acid at 70°C and then cleaned in Piranha solution (3:1) to get rid of residuals and contaminations from the etch process. To decrease surface fluorescence, a final soft-ICP step (30 W, 30 sccm, 10 mTorr) in SF$_6$ plasma was performed to remove remaining residuals and some of the nickel related defects.

To determine the etch angle of the angle etch, the nanobeam cross-section was imaged with a SEM. This etch angle was fed to the FDTD simulations to determine the optimal waveguide geometry and waveguide-fiber efficiency.

## 5.3 Fiber Production

For the waveguide-fiber interface, optical 1060XP fibers from *Diamond SA* were stripped down to their claddings and etched into conical shape by usage of 49 % aqueous HF solution. The pulling speed was held constant to obtain a uniform fiber angle. The HF solution was covered by an O-Xylene layer to interrupt the etching process and prevent post-etching[22]. Finally, the fiber was cleaned with acetone and isopropanol to remove all residuals from the oil.

## 5.4 Spin manipulation

For the ODMR measurements, a 20 µm thick copper wire was carefully stretched across the sample parallel and about 20 µm apart from the waveguides. Note that incautious stretching can easily damage the nanostructures. The microwaves were generated by a SMIQ 03B (*Rhode & Schwarz*) and amplified by a high-power amplifier LZY-22+ (*Mini-Circuits*).

For the Rabi measurement, a ZASWA-2-50DR (*Mini-Circuits*) was used to generate MW and laser pulses. A Pulse Streamer (*Swabian Instruments*) steered the sequence.

## 5.5 Simulation of strain and Stark-shift parameters

We apply density functional theory (DFT) calculations using the Heyd-Scuseria-Ernzerhof (HSE06) functional[41] to model the strain and electric field susceptibility of the zero-field splitting (ZFS) in the quartet spin ground state of the negatively charged silicon vacancy defect at the k-site of 4H-SiC. The calculations are performed in the plane wave based Vienna Ab initio Simulation Package (VASP)[42–45]. The plane wave cutoff energy is set to 420 eV. The contribution of the core electrons is treated in the PAW method[46]. We model the structure in a 576-atom supercell and relax the atomic positions in the unperturbed ground state and under the effect of strain until the forces reach 0.01 eV/Å threshold. Owing to the large supercell, sampling only the $\Gamma$-point in the k-space is sufficient. The ZFS parameters are calculated within VASP as implemented by Martijn Marsman based on Ref. [47]. The effect of electric field on the wavefunction is calculated using the self-consistent response method by adding a homogeneous electric force field to the electric enthalpy functional. The atomic positions remain fixed during these calculations.

## 6. Acknowledgments

We acknowledge Prof. Marina Radulaski, Dr. Petr Siyushev, Dr. Wolfgang Knolle, Dr.-Ing. Patrick Berwian, Erik Hesselmeier, Di Liu, Marion Hagel and Arnold Weible for experimental help and fruitful discussions. F.K., J.U.H., and J.W. acknowledge support from the European Commission through the QuantERA project InQuRe (Grant agreements No. 731473, and 101017733). F.K. and J.W. acknowledge the German ministry of education and research for the project InQuRe (BMBF, Grant agreement No. 16KIS1639K), the European Commission for the Quantum Technology Flagship project QIA (Grant agreements No. 101080128, and 101102140), the German ministry of education and research for the project QR.X (BMBF, Grant agreement No. 16KISQ013) and Baden-Württemberg Stiftung for the project SPOC (Grant agreement No. QT-6). J.W. also acknowledges support for the project Spinning (BMBF, Grant agreement No. 13N16219), DFG for GRK2642 (431314977), and DFG for INST (41/1109-1 FUGG). F.K. acknowledges funding by the Luxembourg National Research Fund (FNR) (project: 17792569). F.D.-M. and F.K. acknowledge support from LIST for an internal PhD thesis grant. J.U.H. further acknowledges support from the Swedish Research Council under VR Grant No. 2020-05444 and Knut and Alice Wallenberg Foundation (Grant No. KAW 2018.0071). The simulations have been partially performed using the resources provided by the Hungarian Governmental Information Technology Development Agency (KIFÜ). A.G. acknowledges the National Research, Development, and Innovation Office of Hungary (Grant No. KKP129866) for the National Excellence Program of Quantum-Coherent Materials Project, the Quantum Information National Laboratory (Grant No. 2022-2.1.1-NL-2022-00004) supported by the Cultural and Innovation Ministry of Innovation of Hungary, and the European Commission for the projects QuMicro (Grant No. 101046911) and SPINUS (Grant No. 101135699). J.H.S. acknowledges financial support from the DFG priority program SPP 2244.

# Supplementary Materials for "Precise characterization of a silicon carbide waveguide fiber interface"

Marcel Krumrein[1], Raphael Nold[1], Flavie Davidson-Marquis[2,3], Arthur Bourama[1], Lukas Niechziol[1], Timo Steidl[1], Ruoming Peng[1], Jonathan Körber[1], Rainer Stöhr[1], Nils Gross[4], Jurgen Smet[4], Jawad Ul-Hassan[5], Péter Udvarhelyi[6,7,8], Adam Gali[6,7,8], Florian Kaiser[1,2,3], and Jörg Wrachtrup[1,9]

1    3rd Institute of Physics, IQST, and Research Centre SCoPE, University of Stuttgart, Stuttgart, Germany.

2    Materials Research and Technology (MRT) Department, Luxembourg Institute of Science and Technology, Belvaux, Luxembourg.

3    Department of Physics and Materials Science, University of Luxembourg, Belvaux, Luxembourg.

4    Solid State Nanophysics, Max Planck Institute for Solid State Research, Stuttgart, Germany.

5    Department of Physics, Chemistry and Biology, Linköping University, Linköping, Sweden.

6    Wigner Research Centre for Physics, Budapest, Hungary.

7    Institute of Physics, Department of Atomic Physics, Budapest University of Technology and Economics, Budapest, Hungary.

8    MTA-WFK Lendület "Momentum" Semiconductor Nanostructres Research Group.

9    Max Planck Institute for Solid State Research, Stuttgart, Germany.

## 1. Simulations of optimal waveguide geometry

Finite Difference Time Domain (FDTD) simulations were performed by *Lumerical* to evaluate the optimal waveguide width and emitter position inside of the nanobeam. The waveguide cross-section is assumed to be triangular due the etching technique with undercut using a Faraday cage. The half-opening angle at the apex of the waveguide is fixed at $\gamma \approx 36°$ in the following simulations. To emulate the emitter, a dipole source was placed perpendicular to the nanobeam propagation direction and perpendicular to the waveguide top surface. This fits to the emitter orientation in the fabricated waveguide sample.

First, the dependence on the waveguide width of the transmission into the fundamental mode was calculated. The results are presented in fig. S1a. Note that only the transmission into one waveguide direction is considered here. The values for maximum transmission and optimal waveguide width for different wavelengths are summarized in table S1. For our measurements, we used a waveguide width of 490 nm, because when we designed the entire experiment, we assumed to have a half-opening angle at the apex of the nanobeam of 45°. For this angle, the maximum transmission at a wavelength of 960 nm indeed occurs for a width of 490 nm.

Additionally, the optimal emitter position inside the cross-sectional plane for a waveguide width of 490 nm was determined. For this, the dipole source was placed at different coordinates inside the waveguide as seen in fig. S2b. The highest waveguide coupling can be achieved when the emitter is placed 95 nm below the top surface and in the central position along the *x*-direction. A deviation of 50 nm from this position in *y*-direction leads to a reduction of the coupling efficiency by about 15 % and in *x*-direction by about 10%.

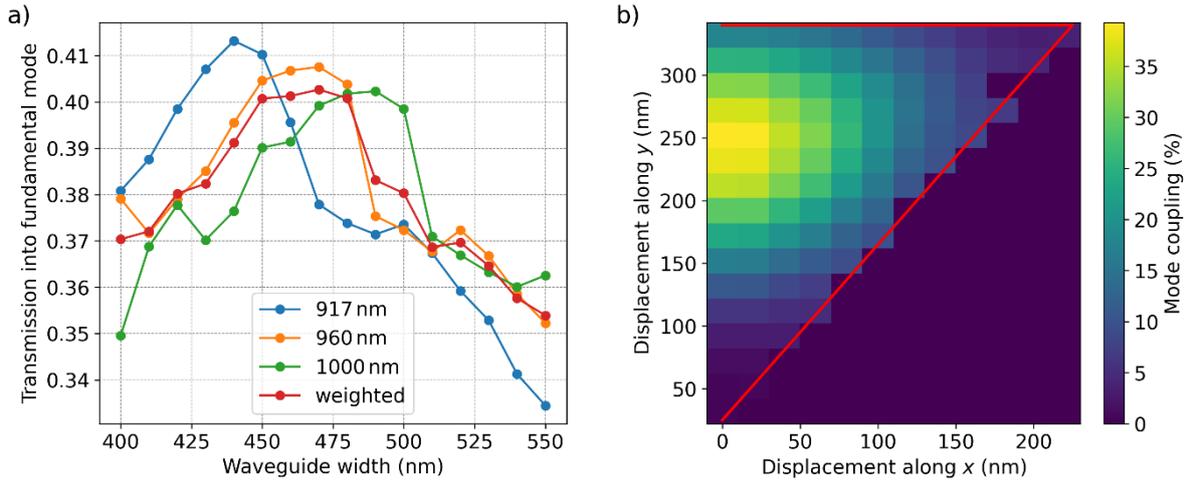

**Figure S1:** (a) Simulation of the transmission into the fundamental mode dependent on the waveguide width for different wavelengths (blue: ZPL, orange: maximum of PSB, green: high-energetic vibrational mode). All three wavelengths were weighted according to the V2 spectrum (8 %: 917 nm, 62 %: 960 nm, 30 % 1000 nm) resulting in the red dots. For widths of up to 580 nm, the transmission is fully single mode with no higher order modes. Note that only the transmission into one waveguide direction is considered here. (b) Transmission into fundamental mode as a function of the dipole position in the cross-sectional plane for a waveguide width of 490 nm. The optimal position is 95 nm below the top surface in the central $x$-position ($x$=0). For negative $x$-values, the mode coupling is mirrored at $x$=0. The red line marks the contour of the nanobeam cross-section.

| Wavelength | Maximum transmission | Waveguide width |
|---|---|---|
| **ZPL at 917 nm** | 41.32 % | 440 nm |
| **PSB at 960 nm** | 40.76 % | 470 nm |
| **PSB at 1000 nm** | 40.23 % | 490 nm |
| **Weighted** | ~40.2 % | 450 − 480 nm |

**Table S1:** Maximum transmission into the fundamental mode for different wavelengths and their corresponding waveguide widths.

## 2. Simulations of waveguide-fiber interface

FDTD simulations of the efficiency of the waveguide-fiber interface for different fiber geometries were performed with the FDTD software package from *Lumerical*. First, the efficiency $\eta$ was determined as a function of the taper-fiber overlap L for various full-opening angles $\beta$ of the tapered fiber (fig. S2a). For $\beta \leq 3°$, the transmission curves as a function of the overlap exhibits a peak. For our fabricated fibers with $\beta = 1.95°$, a theoretical efficiency close to unity can be achieved.

Second, the effect of a broken fiber tip was investigated (fig. S2b). For a tip radius smaller than 200 nm, efficiencies close to unity can be achieved for some range of overlap lengths. As mentioned in the main text, a tip radius of 100 nm is realistic when the fiber is handled carefully. The simulations demonstrate that the overlap range where maximum $\eta$ is achieved gets smaller for larger tip radii. In real experiments, it is therefore possible to estimate the tip radius from the width of the plateau in the efficiency. The simulations also show that both directions of the interface behave in an equal fashion (see data points connected with dashed and solid lines for opposite transmission direction). This is supported by our measurements as well (fig. 2e in main text).

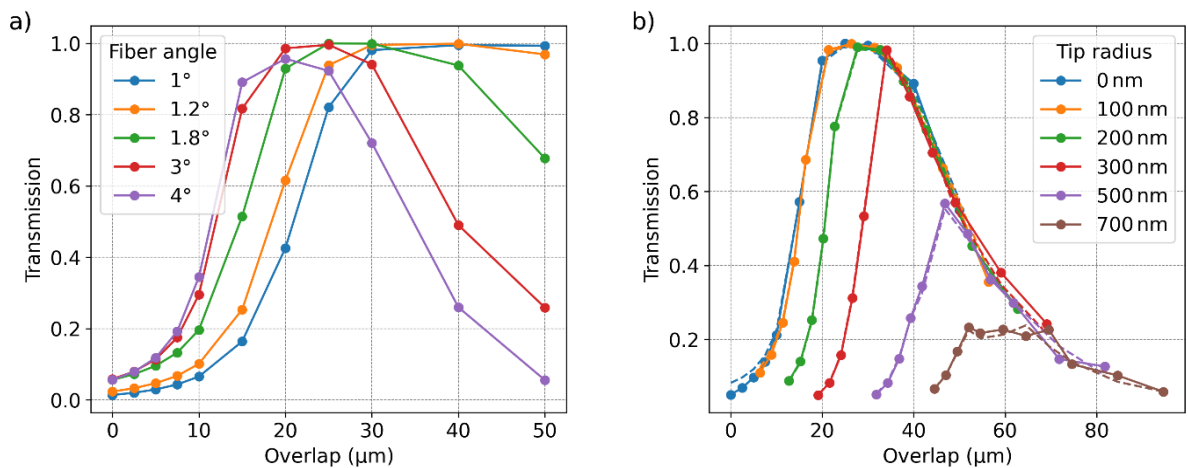

**Figure S2:** Efficiency of the waveguide-fiber interface for different fiber angles and broken tip radii. (a) Transmission of the fundamental mode from the waveguide into the fiber as a function of the taper-fiber overlap L. The full-opening angle of the fiber is varied here. (b) Transmission of the fundamental mode as a function of the taper-fiber interface for different broken tip radii. The data points connected by a solid line are for transmission from the waveguide into the fiber. Data points connected by a dashed line are the opposite direction, i.e. from the fiber into the waveguide.

Fig. S3 summarizes the results of simulations on the wavelength dependence of the transmission across this waveguide and tapered fiber interface. The simulations cover the wavelength range from 917 nm up to 1063 nm.

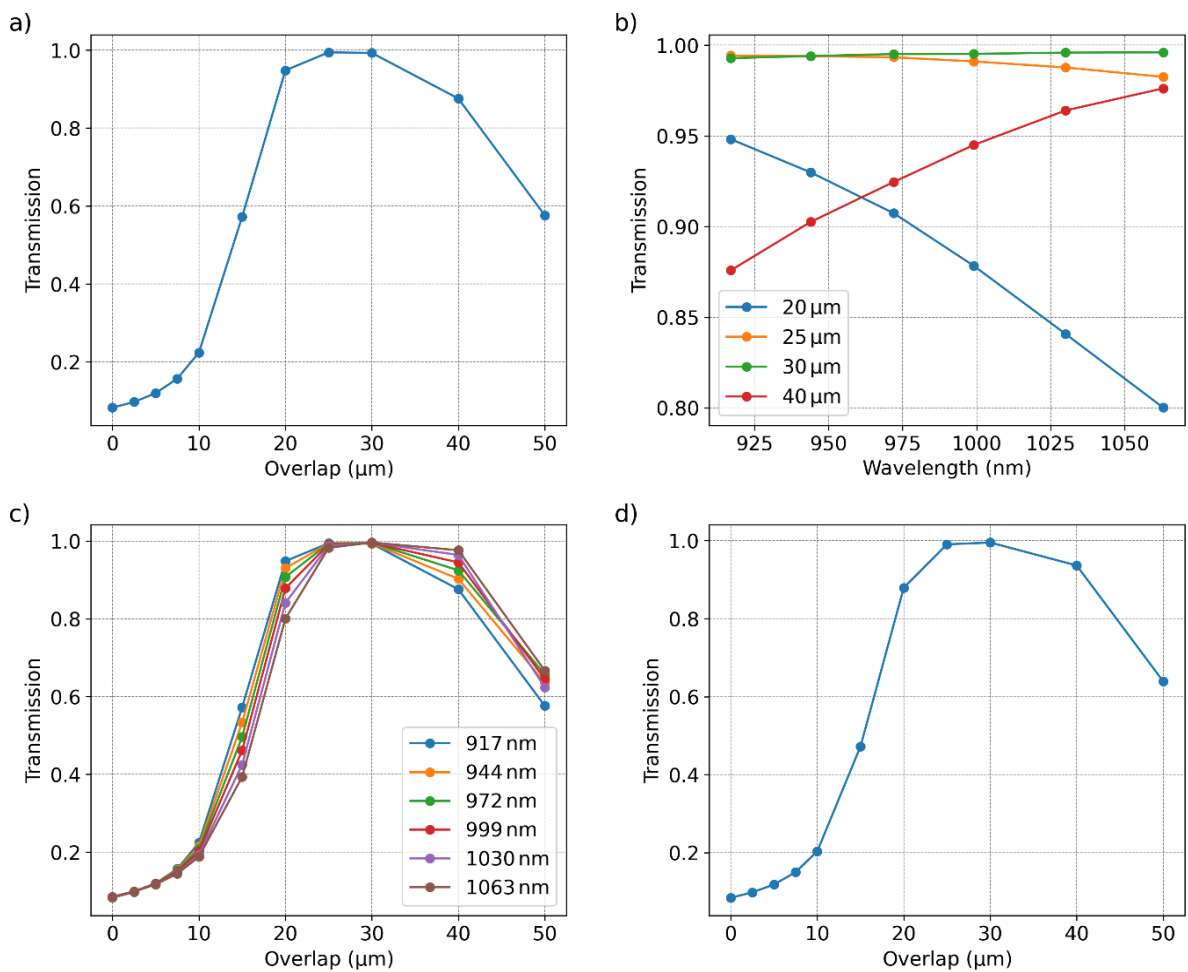

**Figure S3:** Coupling efficiency from fiber to waveguide for different wavelengths. (a) Transmission at the ZPL wavelength of 917 nm as a function of the overlap length L. (b) Efficiency for four different overlap lengths and varying wavelength. (c) Efficiency for six different wavelengths as a function of the overlap length L. (d) Transmission for light with wavelength contributions weighted according to the spectrum of a V2 color center (8 %: 917 nm, 62 %: 960 nm, 30 % 1000 nm).

## 3. Measurement of the fiber angle

To estimate the full-opening angle of the tapered optical fibers, SEM imaging was applied. In those images, markers sitting on the edges were set manually and then fitted by straight lines. From their slope, the angle was determined. First, we tried to apply mathematical algorithms to automatically detect the edges of the fiber. But those algorithms were not reliable enough because the contrast at the edges was too low. Hence, a manual analysis was chosen.

## 4. Nickel related defects

The emitter density inside of the waveguide is much higher than expected. To check whether the density of defects only increased in the nanobeam area as a result of the various processing steps or also in the surrounding substrate, confocal microscopy measurements were performed in both areas with a standard home-built confocal microscope using an immersion-oil objective with NA of 1.2. The images are displayed in fig S4. Next to the waveguide (left panel), the emitter density was identical to the emitter density recorded on an unprocessed sample substrate (right panel) that was grown and irradiated at the same time with the exact same parameters. Hence, only the emitter density in the nanobeam is affected by the processing steps.

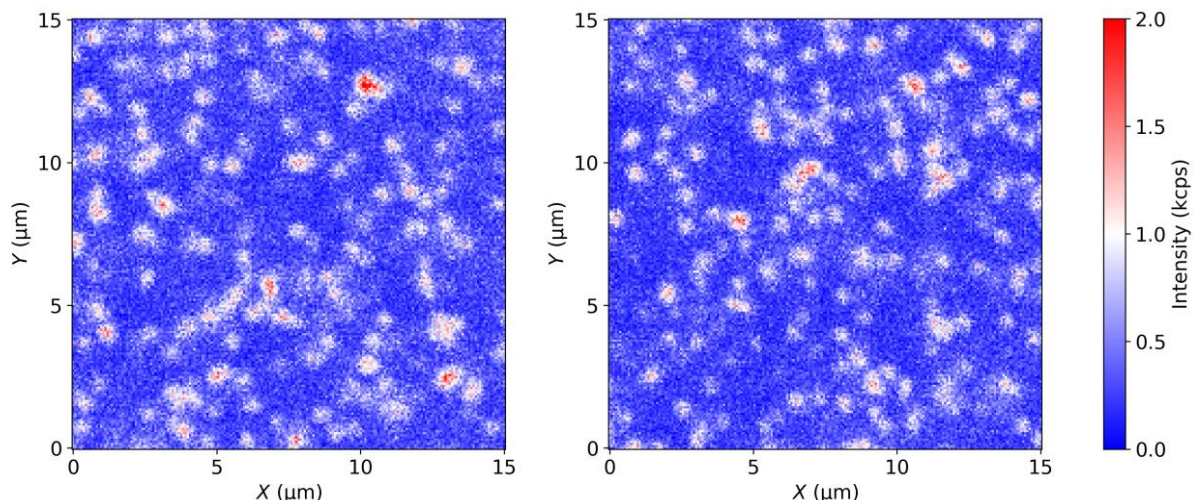

**Figure S4:** Confocal microscopy scan (left) on an area next to the waveguide in the substrate and (right) on a reference sample substrate that underwent the exact same growth and implantation parameters but no processing to create waveguide beams. For detection, avalanche photodiodes (APDs) were used.

In order to understand the origin of the increased emitter density in the nanobeam area, a detailed material study on the formation of those defects was performed. For this purpose, a bare silicon carbide sample was taken, evaporated with a nickel film of 2 nm thickness, followed by several different annealing and etching steps. After each step, a confocal microscopy scan was taken to see the influence of the previous fabrication step. The entire process is summarized in fig. S5.

Both, the bare SiC sample, and the metal evaporated sample do not show any fluorescence. Subsequently, the metal coated sample was annealed at 600°C for 30 minutes. This step created fluorescent emitters. Another sample being evaporated by the same amount of nickel was not annealed but etched for a few nanometers instead. Here, the same formation of defects is visible. Combining the annealing and etching procedures created a very high density of defects with a brightness comparable to that of V2 centers. Hence, by introducing energy into the nickel mask, either by annealing, etching, or both, some nickel atoms diffuse into the silicon carbide and form nickel defects. Accordingly, we assume those emitters to be located close to the surface. A subsequent etching step removing about 20 nm of the surface indeed reduces the density of those nickel related defects significantly. By further removing silicon carbide from the surface, the density of the emitters continues to drop but at a much slower rate. After etching 2 μm, almost all emitters have vanished.

We conclude that most of the emitters in the waveguides are created during the fabrication procedure and are probably related to diffusion of nickel atoms from the evaporated nickel film into the supporting substrate. By removing a few nanometers of the sample surface, a lot of those emitters can already be removed. To get rid of all defects, a deeper etch is required. Unfortunately, this is no solution to reduce the nickel related emitter density in waveguides.

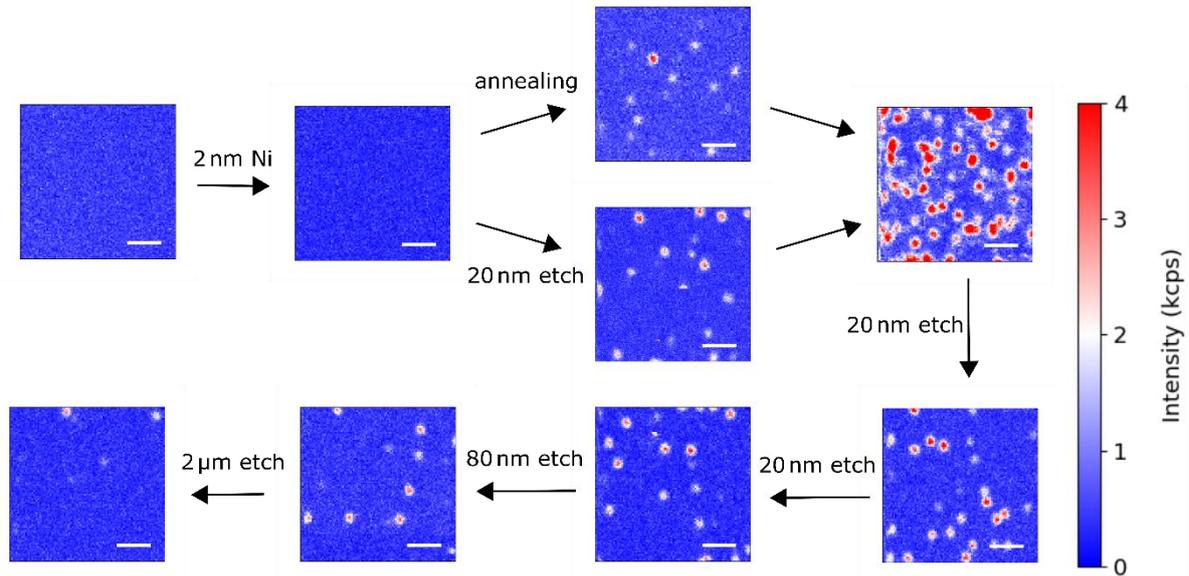

**Figure S5:** Formation and removal of nickel related surface defects in silicon carbide. The white scalebar corresponds to 2 µm in all confocal microscopy images. For detection, superconducting nanowire single-photon detectors were used.

## 5. Raman spectroscopy

To estimate the strain inside the nanobeams, we performed Raman spectroscopy. For the measurements, we used a laser beam with a wavelength of 633 nm and a power of 6 mW. Raman maps were recorded by moving the sample with piezo controllers. The incident light was aligned parallel to the c-axis of the crystal (perpendicular to the surface). Fig. S6 shows a typical spectrum with the Raman-active phonon modes E1(TO), E2(TO) and A1(LO). The peak positions were fitted with a Voigt function and their shift $\Delta \upsilon$ was determined by with a Raman spectrum recorded far away from the waveguide where we expect unstrained bulk material. The Raman shift of various modes depends on both the parallel and perpendicular stress $\sigma_{\mathrm{para}}$ and $\sigma_{\mathrm{perp}}$ according to the following equations[1]:

$$\Delta \upsilon_{E_1} = 2a_{E_1}\sigma_{\mathrm{perp}} + b_{E_1}\sigma_{\mathrm{para}}$$

*(1)*

$$\Delta \upsilon_{E_2} = 2a_{E_2}\sigma_{\mathrm{perp}} + b_{E_2}\sigma_{\mathrm{para}}$$

*(2)*

$$\Delta \upsilon_{A_1} = 2a_{A_1}\sigma_{\mathrm{perp}} + b_{A_1}\sigma_{\mathrm{para}}$$

*(3)*

with the phonon-deformation potential constants $a_{E_1} = -2.06 \ \mathrm{cm}^{-1}/\mathrm{GPa}$, $a_{E_2} = -1.55 \ \mathrm{cm}^{-1}/\mathrm{GPa}$, $b_{E_1} = -0.43 \ \mathrm{cm}^{-1}/\mathrm{GPa}$ and $b_{E_2} = -0.74 \ \mathrm{cm}^{-1}/\mathrm{GPa}$[1]. As these values have not been reported in the literature for the A1(LO) peak yet, we calculated $\sigma_{\mathrm{para}}$ and $\sigma_{\mathrm{perp}}$ from eqs. (3) and (4) and inserted those values into eq. (5) to determine $a_{A_1}$ and $b_{A_1}$. We obtain $a_{A_1} =$ $-1.124 \text{ cm}^{-1}/\text{GPa}$ and $b_{A_1} = -0.651 \text{ cm}^{-1}/\text{GPa}$. To minimize the error in determination of the stress components, we calculated multiple sets of $(\sigma_{\text{para}}, \sigma_{\text{perp}})$ by first using eqs. (1) and (2), then eqs. (1) and (3) and finally eqs. (2) and (3). At the end, we average over all three values. For stress to strain conversion, we use the relations[1]:

$$\varepsilon_{\text{perp}} = \frac{C_{33}\,\sigma_{\text{perp}} - C_{13}\,\sigma_{\text{para}}}{C_{11}C_{33} + C_{12}C_{33} - C_{13}^2},$$

*( 4 )*

and

$$\varepsilon_{\text{para}} = \frac{2C_{13}\,\sigma_{\text{perp}} - (C_{11} + C_{12})\,\sigma_{\text{para}}}{2C_{13}^2 - C_{33}(C_{11} + C_{12})},$$

*( 5 )*

with the elastic stiffness tensor components $C_{11} = 501 \text{ GPa}$, $C_{12} = 111 \text{ GPa}$, $C_{13} = 52 \text{ GPa}$ and $C_{33} = 553 \text{ GPa}$[1]. The resulting strain map for one waveguide is shown in fig. S7.

The parallel strain component $\varepsilon_{\text{para}}$ can take on values up to -0.1 %, and the perpendicular component $\varepsilon_{\text{perp}}$ values up to -0.35 %. Those maximum values match very well with the numbers necessary to explain the observed ODMR shifts (see main text).

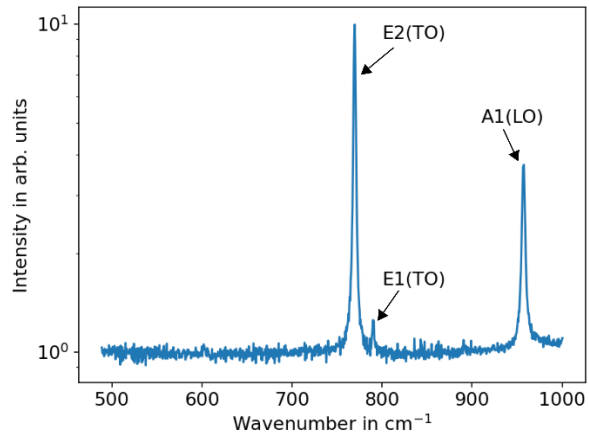

**Figure S6:** Raman spectrum of 4H-SiC at position of unstrained material with a laser at 633 nm and 6 mW. The spectrum contains the Raman-active phonon modes E1(TO), E2(TO) and A1(LO). Cosmic events were excluded from the data.

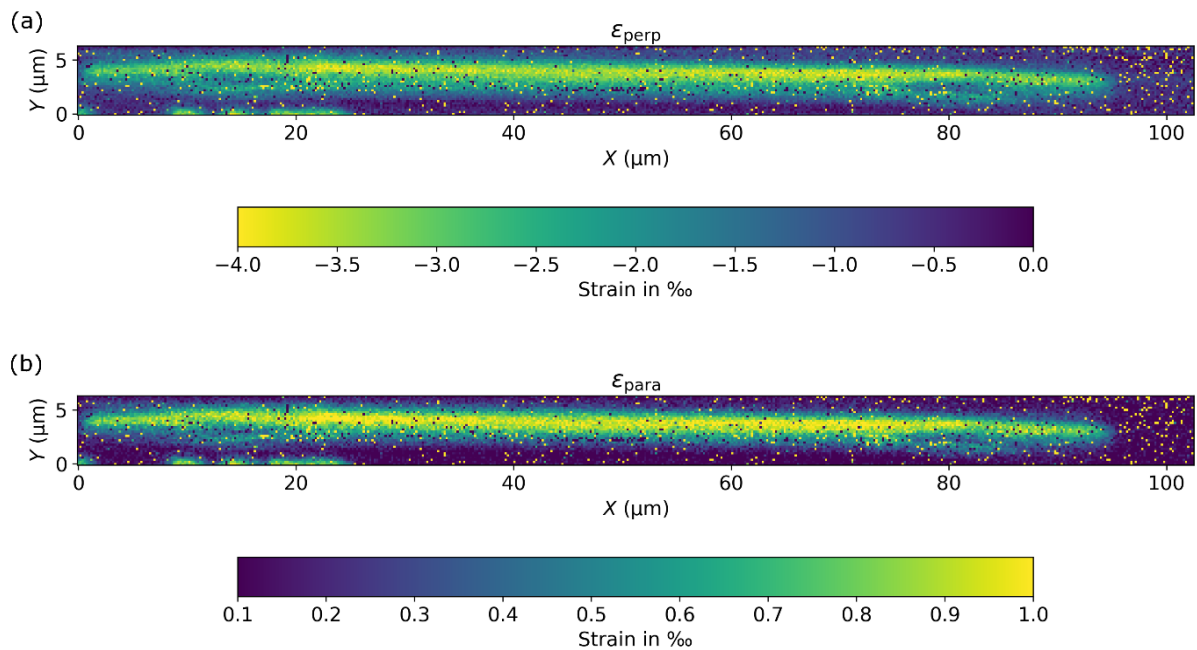

**Figure S7:** Strain map of one waveguide for (a) the parallel strain component $\varepsilon_{\mathrm{para}}$ and (b) the perpendicular strain component $\varepsilon_{\mathrm{perp}}$.

## 6. Full set of characterization measurements for another V2

Another V2 center (emitter #2) was investigated and the auto-correlation measurement, the ODMR spectrum and the saturation study are displayed in fig. S8. The ODMR peak has its maximum at 91.7 MHz indicating that this defect is embedded in an environment with large strain. The ODMR contrast is outstanding for a single V2. The auto-correlation measurement in fig. S8b reveals an antibunching dip of $g^2(0) = 0.466 \pm 0.003$. Accordingly, there is a significant contribution of interfering fluorescence, most likely from a nearby surface defect. Finally, the saturation study in fig. S8c provides a saturation intensity of 224 kcps which is significantly higher than the defect discussed in the main text in fig. 3. This can be explained by poorer antibunching behavior (and, thus, more background emission) and a different position of the defect in the nanobeam cross-section.

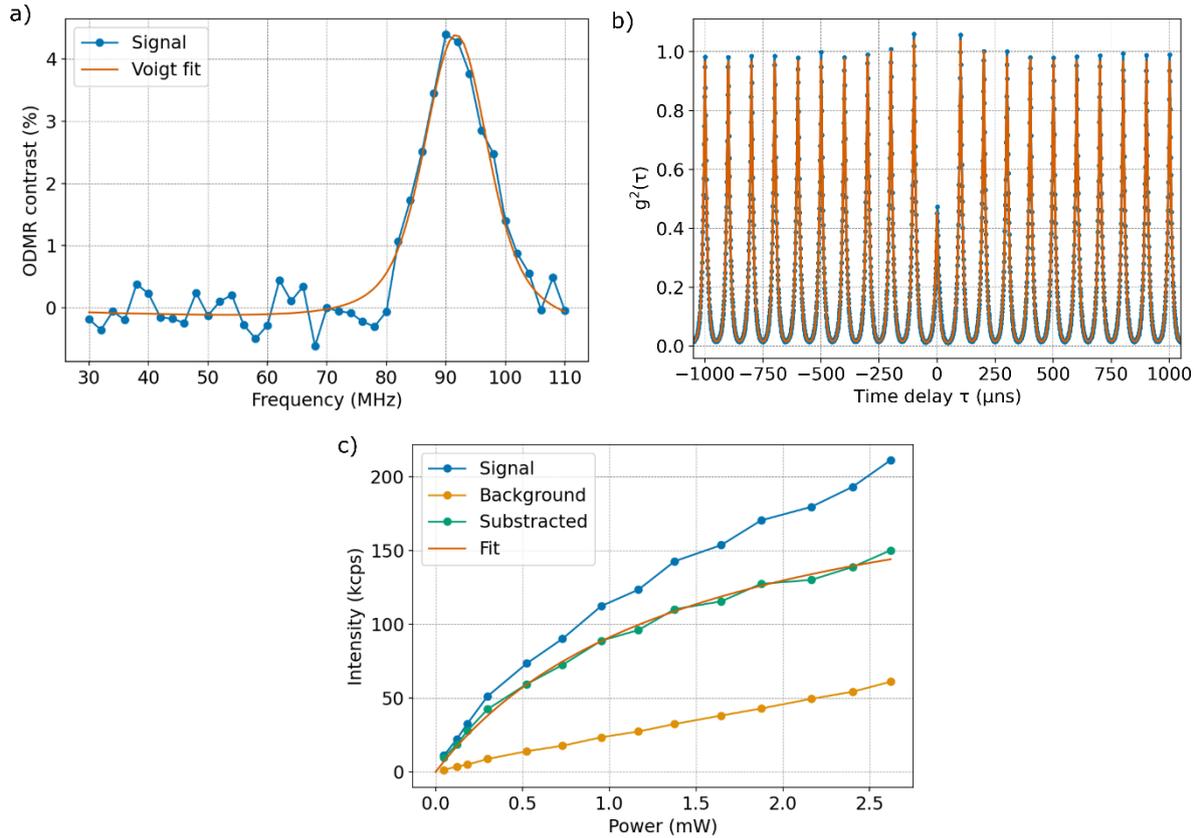

**Figure S8:** Full set of characteristic measurements for a second V2 center. (a) ODMR spectrum with peak maximum at 91.7 MHz and FWHM of 13.0 MHz with an ODMR contrast of >4%. (b) Second order auto-correlation measurement with an antibunching dip of $g^2(0) = 0.466 \pm 0.003$. (c) Fluorescence intensity as a function of the incident power. To isolate the defect emission, the background was recorded and subtracted from the signal. The corrected saturation curve was fitted by eq. (6) resulting in a saturation count rate of 224 kcps.

## 7. Comparison of fluorescence intensity

Estimating the enhancement of the count rate for the waveguide integrated V2 emitters by comparison to bulk measurements is difficult since the count rate of emitters located in bulk material is strongly influenced by the optical components that are used. In particular, the numerical aperture of the objective and the pinhole diameter play an important role and the fluorescence intensity can vary by as much as a factor of two. For collecting the emission of the waveguide integrated defects, no confocal imaging system is used or required and determining an enhancement factor for the emission is of little meaning. Here, we simply compare the brightness of our emitters to those previously reported in literature. Fig. S11 illustrates the saturation count rates for our waveguide embedded emitters and for emitters studied in the literature either in bulk material, in pillars, solid immersion lenses (SILs), or cavities. The numbers behind the symbols refer to the references section at the end of the supplementary material. The waveguide integrated emitters are significantly brighter than emitters in bulk material, SILs and pillars. Only by boosting the ZPL via the Purcell effect (marked as cavity in fig. S11), higher count rates can be achieved.

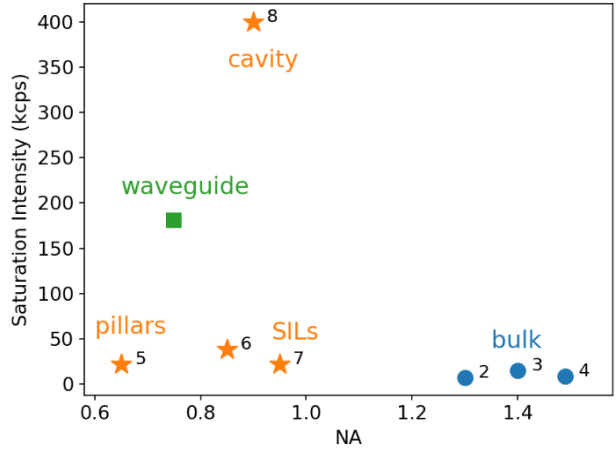

**Figure S9:** Comparison of the fluorescence intensity of the waveguide emitters (green square) with previously reported values for emitters in bulk material (blue circles), and pillars, SILs and cavities (star symbols). The numbers [2–8] behind the symbols refer to the references section at the end of the supplementary material.

## 8. Saturation studies and signal-to-noise of the V2 emitters

For emitter #1 (fig. 3 and S10) and #2 (fig. S8 and S11), the saturation curves were measured for various repetition rates of the laser pulses. For continuous wave excitation, the saturation intensity is described by:

$$I_{\mathrm{cw}}(P) = \frac{I_s \cdot P}{P + P_s},$$

$$(6)$$

with the saturation Power $P_s$ and the saturation intensity $I_s$. For pulsed excitation, the curve is better described by an exponential equation:

$$I_{\mathrm{pulsed}}(P) = I_s \cdot \left[1 - \exp\left(-\frac{P}{P_s}\right)\right].$$

$$(7)$$

For small repetition rates below 20 MHz, the emitter is already decayed back into its ground state either via direct emission or via the metastable state before a second excitation pulse arrives. In this case, the fluorescence intensity is described best by eq. (7). For larger repetition rates above 20 MHz, the emitter is not fully decayed into its ground state and, thus, the excitation is limited by the lifetime of the metastable state. This case is best described by eq. (6).

In figs. S9 and S10, the saturation curves and SNRs for different repetition rates are shown for emitter #1 and #2, respectively. For powers levels below 1 mW, the SNR for emitter #1 is between 5 and 8 and decreases for higher powers. For defect #2, the SNR lies between 3.5 and 6 for an excitation power below 1 mW. This difference agrees well with the g2 and saturation measurements.

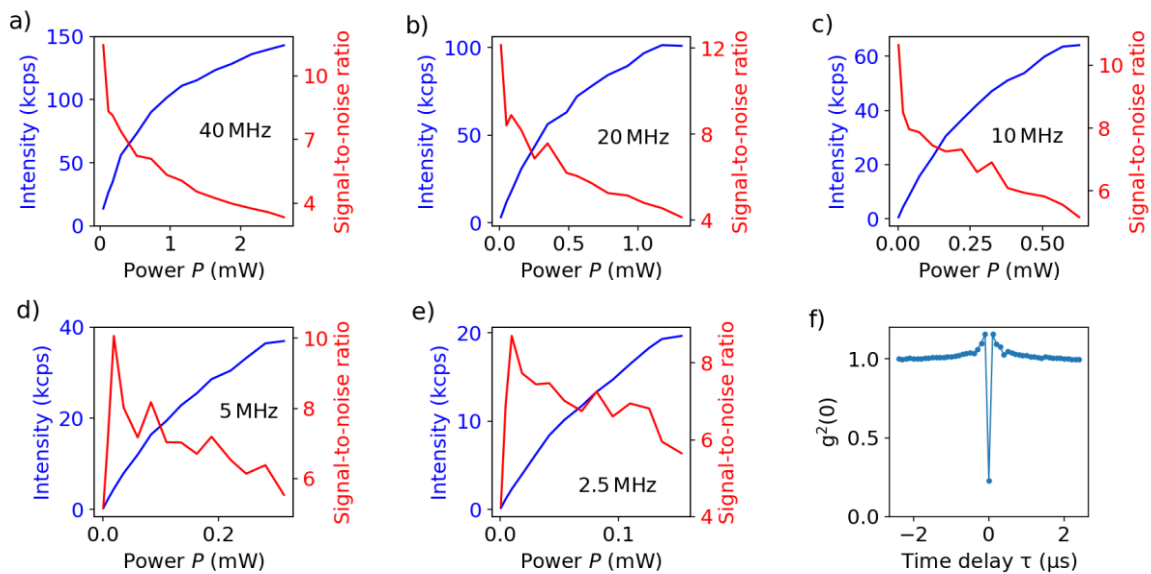

**Figure S10:** Saturation curves and signal-to-noise ratio for emitter #1 for repetitions rates of (a) 40 MHz, (b) 20 MHz, (c) 10 MHz, (d) 5 MHz and (e) 2.5 MHz. In (f), the envelope of the g2 function is displayed when integrating the area below the peaks in fig. 3d (main text).

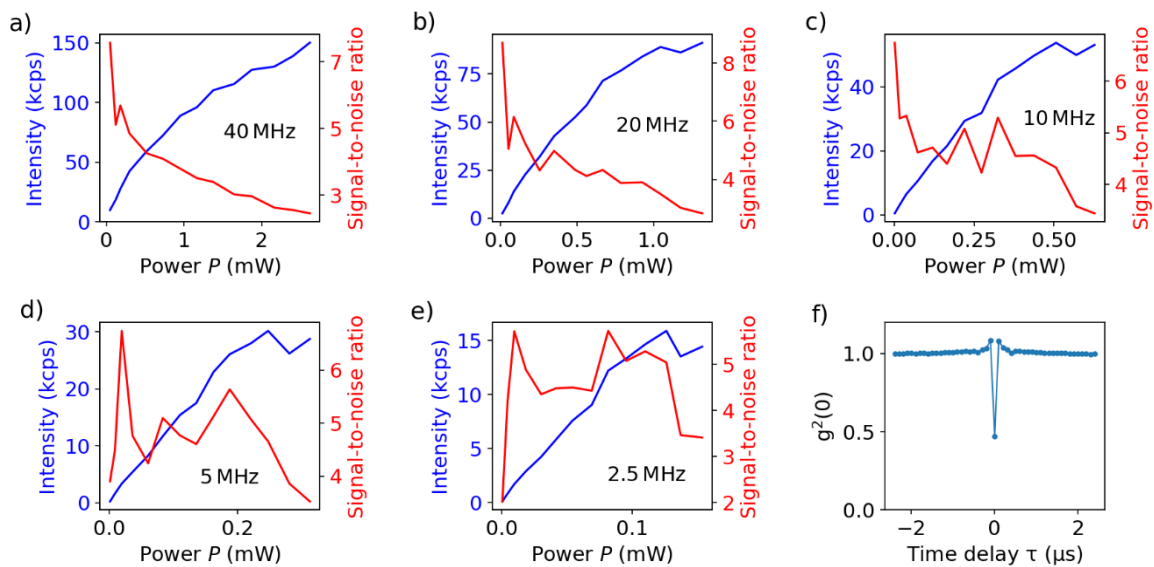

**Figure S11:** Saturation curves and signal-noise ratio for emitter #2 for repetitions rates of (a) 40 MHz, (b) 20 MHz, (c) 10 MHz, (d) 5 MHz and (e) 2.5 MHz. In (f), the envelope of the g2 function is displayed when integrating the area below the peaks in fig. S8b.